\documentclass{aa}

\usepackage{graphics}
\newcommand{\be}{\begin{equation}}
\newcommand{\ee}{\end{equation}}
\newcommand{\bea}{\begin{eqnarray}}
\newcommand{\eea}{\end{eqnarray}}

\begin{document}

\thesaurus{12(12.03.1; 02.07.2)}

\title{Detectability of the primordial origin of the gravitational wave 
background in the Universe}

\author{J. Lesgourgues\inst{1}
\and
D. Polarski\inst{2,3}
\and
S. Prunet\inst{4,5}
\thanks{send offprint requests to prunet@cita.utoronto.ca}
\and 
A. A. Starobinsky\inst{6}}

\institute{SISSA-ISAS, Via Beirut 4, I-34014 Trieste, Italy
\and
LMPT, UPRES-A 6083 CNRS, Universit\'e de Tours, Parc de Grandmont, F-37200 Tours, France
\and
DARC, Obs. de Paris-Meudon, 92195 Meudon Cedex, France
\and
IAP, CNRS, 98bis Boulevard Arago, 75014 Paris, France
\and
CITA, 60 St George Street, Toronto, Ontario M5S 3H8, Canada 
\and 
Landau Inst. for Theoretical Physics, Kosygina St. 2, 
Moscow 117334, Russia}


\titlerunning{Detectability of the primordial origin of GW ...}

\maketitle

\begin{abstract}
The appearance of peaks in the Fourier power spectra of various primordial
fluctuations is a generic prediction of the inflationary scenario.
We investigate whether future experiments, in particular the satellite 
experiment PLANCK, will be able to detect the possible appearance of 
these peaks in the B-mode polarization multipole power spectrum. 
This would yield a conclusive proof of the presence of a primordial 
background of gravitational waves.
\end{abstract}

\section{Introduction}
Early Universe cosmology is reaching a stage where theories put forward 
for the generation of primordial fluctuations can be severely constrained 
by observations. It is already the case with present day observations and 
this will be even more so in the near future due in particular to the 
Cosmic Microwave Background (CMB) anisotropy measurements with unprecedented 
resolution by the satellites MAP (NASA) and PLANCK (ESA). At present, 
only inflationary scenarios seem capable to explain the existing bulk of 
data, in particular the acoustic (Doppler) peak in the CMB, and 
one hopes that the increasing amount of observations will finally lead us 
to the ``right'' inflationary model or at least restrict the remaining viable 
models to only a small number.

We would like here to deal with a generic aspect, one that is common to 
all inflationary models, 
%
%
namely the time coherence of the cosmological 
perturbations. All inflationary scenarios have in common an accelerated stage 
of expansion during which fluctuations are generated on super-horizon scales, 
i.e. with wavelength larger than the Hubble radius. The fluctuations 
responsible for the CMB fluctuations, whether temperature fluctuations 
or polarization, though they originate from vacuum quantum fluctuations, 
were for a long time on ``super-horizon'' scales and this is why they appear 
to us as classical fluctuations with random amplitude and fixed temporal 
phase. In other words, soon after the end of inflation, cosmological 
perturbations appear to consist of only the growing, or quasi-isotropic, modes 
with an excellent accuracy.
Remarkably enough, this coherence has a very distinct observational 
signature resulting in periodic acoustic peaks in the CMB temperature 
anisotropy multipoles $C^{\rm S}_{\rm l}$ and also in the corresponding multipoles 
of the CMB 
polarization. Hence, the detection of these periodic peaks would be a 
dramatic confirmation of their primordial origin. 

As well known, the generation of a gravitational wave (GW) background on a 
vast range of 
frequencies is also an important prediction of inflationary models (first 
quantitatively calculated in Starobinsky 1979), one 
that could constitute, if observed, a crucial experimental confirmation of 
these scenarios. In addition, what was said above concerning the time 
coherence 
of the fluctuations is equally valid for the primordial scalar fluctuations 
as well as for the primordial tensorial fluctuations, or primordial GW 
background. For them too, their primordial origin will uncover itself 
in the presence 
of a periodic structure in the multipole power spectrum which we call 
primordial peaks. Clearly, they are much more difficult to track than 
acoustic peaks produced by scalar (energy density) fluctuations.  
Note that these primordial peaks are periodic, with a periodicity (Polarski 
\& Starobinsky 1996)
\be
\Delta l= \pi \Bigl ( \frac{\eta_0}{\eta_{\rm rec}}-1 \Bigr )~,
\label{Del}
\ee 
which is approximately half the spacing between
primordial acoustic peaks produced by scalar fluctuations (due to the
difference between the light velocity which is relevant for (1) and
the sound velocity in the baryon-photon plasma at recombination which
enters into the corresponding expression for the spacing between acoustic
peaks). Note that, strictly speaking, Eq.(1) becomes exact for 
$l\to\infty$ only. However (see Fig.~\ref{fig1}), it turns out that Eq.(1) is
already a good approximation for the spacing between the first and second
peaks. In Eq.(\ref{Del}), $\eta\equiv \int^t \frac{dt'}{a(t')}$ and $\eta_0$, 
resp. $\eta_{\rm rec}$ are evaluated today, resp. at recombination.

Of course, the detection of these peaks is much more complicated than 
the discovery of a long-wave GW background in the Universe through the
B-mode polarization of the CMB, though such a discovery would represent
a great achievement in itself (for its prospects see, e.g., Kamionkowsky
\& Kosowsky 1998). However,
the significantly smaller effect which we consider in this paper 
- the existence
of multiple primordial peaks in the angular spectrum of the B-mode CMB 
polarization -  is fundamental and remarkable enough to justify
hard efforts to detect it for two reasons. The first reason, explained above, 
concerns the primordial origin of the GW background; the
second one is related to the use of 
Eq. (1) in order to determine fundamental cosmological parameters.

The discovery of the (asymptotic) periodicity of the 
$\Delta T/T$ peaks produced by
a primordial GW background will immediately give us an unbiased value
of one of the most important parameters: the ratio of the present conformal 
time to the recombination conformal time $\eta_0/\eta_{\rm rec}$. Furthermore,
by combining this result with the periodicity scale of acoustic peaks
in the CMB anisotropy and the E-mode of the CMB polarization at $l>200$ (which
is much easier to measure) we can directly find the value of the sound
velocity $c_{\rm s}$ in the cosmic photon-baryon plasma at the moment of 
recombination. This, in turn, leads to a new way of determining the present
baryon density $n_{\rm B}$ which is free of "cosmic confusion".

Actually the observation of the peaks in the multipoles $C^T_{\rm l}$
due to the primordial GW is a hopeless experimental challenge 
with the  presently existing technology.
On the other hand, the observation of this coherence in a direct detection 
experiment of the primordial GW background is even worse: it would require 
a resolution in frequency $\Delta \nu\approx 10^{-18}$Hz (as briefly 
mentioned in Polarski \& Starobinsky 1996, p.389), something that is 
clearly impossible to achieve (see also Allen, Flanagan \& Papa 2000 
for a recent careful investigation).

A better prospect for the detection of these peaks might perhaps be
offered by the measurement of the CMB polarization as scheduled by
PLANCK.  We expect the CMB to be also polarized and important physical
information could be extracted from it. In particular, the scalar
fluctuations will not contribute to the so-called B-mode polarization
(Kamionkowski et al. 1997a; Seljak \& Zaldarriaga
1997), therefore the latter bears the imprint of the primordial GW
only.
%
%
Hence, 
%
%
CMB polarization measurements might enable us to show the presence of a GW 
background of primordial origin. 
It is the aim of this letter to investigate whether the sensitivity 
of PLANCK is sufficient for this purpose. We will do this using a concrete, 
viable model (Lesgourgues et al. 1999a, 1999b) in which 
the generated GW background can be fairly high, with 
$C^{(T)}_{10}\leq C^{(S)}_{10}$ (note that here, $C^{(T)}_{\rm l}$, resp. 
$C^{(S)}_{\rm l}$, stands for the temperature anisotropy 
multipoles produced by tensorial, resp. scalar, perturbations).

\section{The model and the induced polarization}

The primordial GW produced during the inflationary stage originate from 
vacuum fluctuations of the quantized tensorial metric perturbations. Each 
polarization state $\lambda$ -- where $\lambda=\times,+$, and the
polarization tensor is normalized to $e_{\rm ij}({\bf k})~e^{\rm ij}({\bf k})=1$ -- has an 
amplitude $h_{\lambda}$ (in Fourier space) given by
\begin{equation}
h_{\lambda}=\sqrt{32\pi G}~\phi_{\lambda}
\end{equation}
where $\phi_{\lambda}$ corresponds to a real massless scalar field. The 
production of a GW background is a generic feature of all inflationary models.

Let us briefly describe the BSI (Broken Scale Invariant) inflationary model 
used here. The power spectrum of this model has a characteristic scale 
which is due to a rapid change in slope of the inflaton potential $V(\varphi)$
from $A_+>0$ to $A_->0$ (when $\varphi$ decreases) 
in some neighbourhood $\Delta\varphi$ of $\varphi_0$ (Starobinsky 1992).
As a consequence, one of the two slow-roll conditions is violated and this 
is why the scalar perturbation spectrum $k^3\Phi^2(k)$ is non-flat around
the scale $k_0=a(t_{\rm k_0}) H_{\rm k_0}$, which becomes larger than the 
Hubble radius when $\varphi(t_{\rm k_0})=\varphi_0$ ($H\equiv \dot{a}/a$ is 
the Hubble parameter).
The spectrum can be basically represented as ``step-like'' 
%
%
 while its shape is determined solely by the parameter 
$p\equiv \frac{A_-}{A_+}$ and is independent of the characteristic scale $k_0$.
In particular, an inverted step is obtained for $p<1$.
This model could nicely account for the possible appearance of a spike in 
the matter power spectrum (Einasto et al. 1997).
We will assume that the inflaton potential 
satisfies the slow-roll conditions far from the point $k_0$ and consider 
a particular behaviour of the spectral indices $n_{\rm T}(k)$ and $n_{\rm s}(k)$.
This model was thoroughly investigated previously (Lesgourgues et al. 1999b; 
Lesgourgues et al. 1999; Polarski 1999) and 
it was found 
to be in agreement with observations in the presence of a large cosmological 
constant ($\Omega_{\Lambda}\approx 0.7$, as favoured by recent 
observations). 
We refer the interested reader to the literature for further technical 
details about our model and the possible observational hints in support of 
its BSI spectrum.

Also our model allows 
a high fraction of the temperature anisotropy to originate from tensorial 
fluctuations with $C^{(T)}_{10}\leq C^{(S)}_{10}$. This last property is 
significantly different from scale free single-field slow roll inflation 
for which the height of the Doppler peak precludes a high contribution of 
the GW to $\frac{\Delta T}{T}$ on large angular scales where the power 
spectrum gets normalised. It is this fact which is of interest to us here 
as we may hope that the B-polarization is large enough for our purposes. 

We introduce now the polarization tensor and the multipole power spectra 
needed besides $C^T_{\rm l}$, where 
\be
\langle a^{T*}_{\rm lm}a^T_{\rm l'm'} \rangle \equiv C^T_{\rm l}~\delta_{\rm ll'}~\delta_{\rm mm'} 
\ee
and the coefficients $a^T_{\rm lm}$ are defined through
\be
\frac{\Delta T}{T}= \sum_{\rm l=0}^{\infty}\sum_{\rm m=-l}^{m=l}a^T_{\rm lm}~Y_{\rm lm}~.
\ee  
The symmetric, trace-free polarization tensor $P_{\rm ab}$ can be expanded 
as follows 
\be
\frac{P_{\rm ab}}{T}=\sum_{\rm l=0}^{\infty}\sum_{\rm m=-l}^{m=l}
\Bigl (a^E_{\rm lm}~Y_{\rm lm,ab}^E +a^B_{\rm lm}~Y_{\rm lm,ab}^B \Bigr )~,  
\ee
where $Y_{\rm lm}^{E,B}$ are electric and magnetic type tensor spherical 
harmonics, with parity $(-1)^l~{\rm and}~(-1)^{l+1}$ respectively. A  
description of the CMB requires the three power spectra 
\be
C^T_{\rm l}\equiv \langle |a^T_{\rm lm}|^2 \rangle,~~~~~
C^E_{\rm l}\equiv \langle |a^E_{\rm lm}|^2 \rangle,~~~~~
C^B_{\rm l}\equiv \langle |a^B_{\rm lm}|^2 \rangle~,
\ee
together with the only non vanishing cross correlation function
\be
C^{TE}_{\rm l}\equiv \langle a^{T*}_{\rm lm}a^E_{\rm lm} \rangle~.
\ee
Indeed, because of parity, the cross-correlation functions 
$C^{TB}_{\rm l},~C^{EB}_{\rm l}$ vanish. 
Among the different types of primordial perturbations, only the 
primordial GW can produce B-mode polarization. Hence the latter offers a 
unique opportunity to probe the possible presence of a GW background and in 
particular its primordial origin.  

\section{Statistical analysis}

We want first to investigate whether Planck has the required 
sensitivity in order to see possible small peaks in the power spectrum $C^B_{\rm l}$.
Our method will make use of the Fisher information matrix $F_{\rm ij}$. 
  
Using the CMB Boltzmann code {\sc CMBFAST} (Seljak \& Zaldarriaga 1996), 
we compute the derivative of the $C_{\rm l}$'s with respect to each parameter 
$\theta_{\rm i}$ on which the spectra may depend in a given model. 
The Fisher matrix (Jungman et al. 1996a, 1996b; Tegmark et al. 1997; 
see also Bond et al. 1997;
Copeland et al. 1998; 
Eisenstein et al. 1998; Wang et al. 1999;
Stompor \& Efstathiou 1999) 
is then obtained by
adding the derivatives, weighted by the inverse of the 
covariance matrix of the estimators of the polarized and unpolarized
CMB power spectra for the PLANCK satellite mission ,
$\rm{Cov}(C_{\rm l}^X,C_{\rm l}^Y)$: 
\begin{equation}
F_{\rm ij}=\sum_{\rm l=2}^{+\infty}\sum_{\rm X,Y}
\frac{\partial C_{\rm l}^X}{\partial \theta_{\rm i}}
{\rm Cov}^{-1}\left(C_{\rm l}^X,C_{\rm l}^Y\right)
\frac{\partial C_{\rm l}^Y}{\partial \theta_{\rm j}}~,
\end{equation}
where $\{X,Y\} \in \{T,E,B,TE\}$ (Kamionkowski et al.
1997b; Zaldarriaga et al. 1997;
Prunet et al. 1998a, 2000). 
The Fisher matrix $F_{\rm ij}$\ measures basically the width and the shape of 
the likelihood function around the maximum likelihood point.
Assuming that a fit to the PLANCK data yields a maximum likelihood
for the model under consideration (for which the derivatives were computed),
the $1-\sigma$ error on the parameter $\theta_{\rm i}$, for any unbiased estimator 
of $\theta_{\rm i}$ and however precise the observations may be, satisfies 
\be
\Delta \theta_{\rm i} \geq \sqrt{(F^{-1})_{\rm ii}}~,
\ee
if all the parameters are estimated from the data, and   
\be
\Delta \theta_{\rm i} \geq F_{\rm ii}^{-\frac{1}{2}}~,
\ee
when all other parameters are known.

Each multipole will be measured by Planck with unprecedented precision of the
order of 1\%, thereby allowing for an accurate extraction of the cosmological 
parameters.
Still, one should remember that a given model with its spectra implies a
set of parameters, each having a particular value, which define the model. 
Even though the power spectra $C_{\rm l}^X$ for some given parameter 
combination might be 
measured with very high precision, each parameter separately is usually 
constrained only at the percent level due to the possible degeneracy of 
the spectra with respect to a change in the parameter combination. 
In computing the covariance matrix of the CMB power spectra, we accounted
for the presence of foregrounds (both polarized and unpolarized) in
the measurement of the CMB power spectra, using the method described
in Bouchet et al. 1999 (see also Prunet et al. 
1998a, 2000). 

In order to use this approach we need to quantify the appearance of
peaks with the help of some additional parameter $\theta_{\rm i}\equiv
s$. For this purpose, we adopt the following strategy: we compare the
$C^B_{\rm l}$ curve of our inflationary model where peaks are present with a
smoothed version $C^B_{\rm l,sm}$ which contains no peaks anymore. 
Obviously, we can write  
%
%
\be 
C^B_{\rm l}=C^B_{\rm l,sm}+s(C^B_{\rm l}-C^B_{\rm l,sm})~.  
\ee
Hence, the parameter $s$ enters the Fisher matrix through the quantity
\be 
\frac{\partial C^B_{\rm l}}{\partial s}=C^B_{\rm l}-C^B_{\rm l,sm}. 
\ee 
Note that $s=1$ corresponds to the original model which is assumed to
be the correct one.  We stress that it is perfectly self-consistent to
smooth only the $C^B_{\rm l}$ spectrum since the possible appearance of
peaks in the other spectra is due to the scalar perturbations only.
This is well known for the temperature anisotropy, and it is also true
for the E-mode polarization multipoles $C^E_{\rm l}$.  In summary, what we
really measure with the help of the parameter $s$ is the presence of a
time-coherent GW background, in other words, a GW background which is
of primordial origin.

For completeness, we take also into account the additional information 
provided by the $T, E$ and $TE$ modes: we smooth the tensor contibutions 
$C^{X, (T)}_{\rm l}$, $X \in \{T,E,TE\}$, and calculate 
\be 
\frac{\partial C^X_{\rm l}}{\partial s}=C^{X, (T)}_{\rm l}-C^{X, (T)}_{\rm l,sm}. 
\ee 
We stress that in general, statistical separation of the tensor contribution 
from the scalar contribution requires prior knowledge about the 
underlying theory (which is available here by assumption).
Even so, this will change $F_{\rm ss}$ only by a small amount, 
due to observational uncertainties in tensor-scalar separation, a drawback 
which does not affect the B mode.

\begin{figure}
  \resizebox{\hsize}{!}{\includegraphics{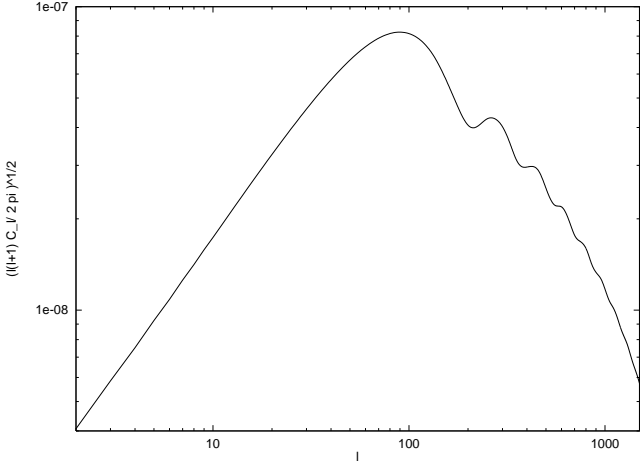}}
  \caption{$C^B_{\rm l}$ curve for the model under consideration. The $C^B_{\rm l,sm}$
  curve is obtain by smoothing this curve for $l \geq 150$.}
  \label{fig1}
\end{figure}

We fix the parameters of our model to $\Omega_{\rm tot}=1,~\Omega_{\Lambda}=
0.65,~\Omega_{\rm b}=0.04,~h=0.6,~p=0.58,~k_0=0.016~h
{\rm Mpc}^{-1},~n_{\rm S}(k < k_0)=1,~n_{\rm T}(k_0)=-0.125$. 
For these parameters, Eq. (1) gives $\Delta l\approx 160$ (assuming
3 kinds of massless or very light neutrinos).
As shown in (Lesgourgues et al. 1999b), this choice is 
consistent with current constraints, despite a
fairly high GW contribution to the CMB temperature anisotropy with
$C_{10}^{(T)} / C_{10}^{(S)} = 0.85$. We find that the $1-\sigma$ error 
$\Delta s$ on the parameter $s$ satisfies 
\be 
\Delta s \geq \sqrt{(F^{-1})_{\rm ss}} = 2.68
\ee 
if all other parameters are extracted from the same data as well, while 
essentially the same result is obtained 
\be 
\Delta s\geq F_{\rm ss}^{-\frac{1}{2}} = 2.63
\ee 
when all other parameters are known. 
This is not surprising since the error in the measurement of this
parameter is dominated by the noise and the foregrounds and not by a
possible degeneracy with the other parameters.  Since in both cases
$\Delta s\geq 1$, Planck clearly does not seem to have the level of
sensitivity required in order to see the primordial peaks in the B-mode
polarization, at least for our model. We recall however that our model
admits a large GW background, in any case substantially larger than in
usual single-field slow-roll inflationary models.  Therefore, a
negative result for this model is almost certain to imply, for the
particular problem under consideration, rather gloomy prospects for
most, if not all, viable inflationary models\footnote{Also, in our
model, it is possible to neglect the gravitational lensing
contamination of the B mode (Zaldarriaga \& Seljak 1998), in contrast
with models with a low tensor contribution. Indeed, in our model,
gravitational lensing generates a B-polarized signal that dominates
the primordial gravitational wave signal for $l>140$. However, we
checked with a specific Fisher matrix analysis that from the
measurement of ${T,E,TE}$ modes alone, the $C^B_{\rm l}$ contamination can
be substracted with 4\% accuracy, and therefore neglected up to
$l=350$, while our result for $F_{\rm ss}$ depends mainly on multipoles
$C^B_{\rm l}$ with $150 < l < 350$.}.

It is interesting to evaluate what is the sensitivity required for 
other future experiments.
If we imagine an idealized experiment, with only one channel,
and no foregrounds contamination at all, we find that
only a sensitivity ten times higher than that achieved by Planck's
best channel will allow a clear detection with $\Delta s \simeq 0.1$.
The assumption of no foregrounds contamination is clearly an idealization if 
we compare the expected level of the dust polarized B-mode power spectrum
(see for instance Prunet et al. 1998b)
to the CMB spectrum shown in Fig.~\ref{fig1}. 

However, the level of contamination is very inhomogeneous on the sky,
and one expects to find some locations where the contamination level by dust
would be at least ten times smaller than the mean level computed for a 
galactic
latitude $\|b\|>20^{\circ}$. Of course, the draw-back of observing a smaller 
part
of the sky is that it increases the sample variance. Indeed, in the 
no-foregrounds
case, the sample variance part of the covariance of the estimator of a 
given B-mode 
multipole $C_{\rm \ell}^B$ is approximately given by 
\be
\Delta C_{\rm \ell}^B/C_{\rm \ell}^B \simeq \sqrt{\frac{2 f_{\rm sky}}{2\ell +1}}
\ee
where $f_{\rm sky}$ is the fraction of the sky covered by the experiment.
However, since we are interested
in multipoles $\ell\ga 250$, a rather small region 
(typically $400\;{\rm deg}^2$)
should be sufficient for this sample variance to be smaller than the noise. 
Thus a dedicated, long-time observation of a particularly clean region of the
sky, like the {\it Polatron} experiment
\footnote{see http://astro.caltech.edu/~lgg/polatron/polatron.html}, 
with possibly a poorer angular resolution than {\it Polatron} but with a 
significant gain in sensitivity, 
should be able to constrain the coherence parameter $s$ to a reasonable
accuracy, especially if we take into account the expected progress in 
bolometer technology.   

In conclusion, it is not unreasonable to expect that in the upcoming decades, 
CMB polarization experiments, in addition to addressing the very existence 
of a cosmic gravitational wave background (which we think will already be 
settled by Planck), will also answer the fundamental question 
concerning the primordial origin of this background. 

\begin{acknowledgements}
A.S. was partially supported by the grant of the Russian Foundation
for Basic Research No. 99-02-16224 and by the Russian Research Project
"Cosmomicrophysics". This paper was finished during his stay at the
Institute of Theoretical Physics, ETH, Z\"urich.
J. Lesgourgues is supported by the European community TMR network grant 
ERBFMRXCT960090. 
\end{acknowledgements}

\end{document}